\documentclass[aps,prl,twocolumn,showpacs]{revtex4}
\usepackage{graphicx}
\usepackage{dcolumn}
\usepackage{bm}
\usepackage{epstopdf}
\usepackage{amstext} 
\usepackage{fixmath} 

\begin{document}

\newcommand{\WT}[1]{\textbf{\textcolor{blue}{(#1 -- WT)}}}
\newcommand{\w}{$\omega$}
\newcommand{\m}{$\mu$}

\title{Comment on Search for Axionlike Dark Matter with a Liquid-State Nuclear Spin Comagnetometer}

\author{E.\,G.~Adelberger}\email{eadelberger@gmail.com}
\affiliation{Center for Experimental Nuclear Physics and Astrophysics, Box 354290,
University of
Washington, Seattle, Washington 98195-4290} 
\author{W.\,A.~Terrano}\email{wterrano@princeton.edu}
\affiliation{Department of Physics, Princeton University, Princeton NJ 08550 USA}

\maketitle

This elegant experiment searched for axionic dark matter "wind"  using two different nuclear magnetometers in the same molecule to cancel magnetic effects. However, the authors used a complicated analysis of the power spectra (described in the Supplementary Material) to obtain the constraint shown in the left panel of their Fig. 3, which is clearly flawed.  Any constraint on oscillations with periods, $\tau$, long compared to the span of the data, $T=2.642\times 10^{6}$~s, cannot be tighter than those on oscillations with periods shorter than $T$ and in fact must be substantially weaker.  This reduction in experimental sensitivity can be seen in other recent works that searched for similar effects\cite{ti:15, he:16, ab:17, te:19}. 

A simple analysis that provides the correct result and quantitative insight is a linear least-squares analysis (LLSA) such as that used in ref.\cite{te:19}.   Here one fits the time-series data and its errors with orthogonal, typically non-linear, basis functions that model the data expected from the physics being probed. In the present case, for each assumed axion mass there will be two such functions of time that contain the non-linear transforms between celestial and laboratory coordinates multiplied by $\sin (2\pi t/\tau)$ or $\cos (2\pi t/\tau)$ where $t$ is the time of the measurement. At each assumed axion frequency one makes a \underline{simultaneous} linear fit of the data to the two basis states, obtaining two fit amplitudes and their errors along with the correlation matrix. Oscillation constraints marginalized over the axion phase are found by combining the quadrature (sine and cosine) amplitudes. When $\xi=\tau/T< 1$ the two quadrature basis functions have virtually similar mean magnitudes and are nearly orthogonal. The resulting constraint of the marginalized amplitude is (barring a real signal and assuming $\delta<<\tau$ where $\delta$ is the duration of an individual measurement) nearly independent of frequency. However when $\xi>1$ the basis states are either highly anticorrelated or now have very different magnitudes. This greatly increases the uncertainties in the marginalized amplitude. As a concrete example, consider the case where the zero of the time is arbitrarily set to zero at the center of the data span. Now the basis functions are proportional to $\sin(2\pi t/\tau)\approx 2\pi t/\tau$ and $\cos(2\pi t/\tau)\approx 1-(2\pi t/\tau)^2/2$ where $-T/2<t<+T/2$. The central value and error of extracted sine amplitude blow up as
$\tau/(\pi T)$, which for oscillations corresponding to the left hand edge of Fig. 3 has a value of 6.0. This argument is quite general and does not depend on a choice of the origin of the time scale. For example, suppose that the origin of time were set so that the phase of the axion signal was $\pi/4$ at the midpoint of the data span. Now neither of the two basis functions tends to zero but the two functions are almost perfectly anticorrelated giving essentially the same blowup factor in the extracted marginalized amplitude. 

We trace the strange long-period behavior in Fig. 3, where the constraint actually becomes tighter as $\pi T/\tau<2.5$, to the discussion in Derivation of The Constraint Level (Eqs. S12-S17) in the supplementary material. These relate the data $\Delta \cal R$
to the axion coupling $g_{aNN}$ and the phase $\phi$ of the axion oscillation, with $g_{aNN}$ proportional to $\Delta {\cal R} /\sin\phi$. The authors account for the effect on $g_{aNN}$ of the unknown phase by averaging $|\sin\phi|$ over the interval for $\phi$ between 0 and $2\pi$. This is incorrect as the constraint is actually proportional to $1/\sin\phi$.  (Note that approaches where the $\phi$ ia a free parameter will not have a pole for any non-zero frequency.)  Assuming that the analysis for signals with $\tau < T$ is correct, the exclusion limit at the left-hand edge of Fig. 3 should be 9.5 times higher than shown. This would be $\sim$ 15\% less constraining than the previous limit set using neutron EDM data. This work was supported in part by National Science Foundation Grants PHY-1305726 and PHY-1607391.

\end{document}